\documentclass[10pt, conference, letterpaper]{IEEEtran}

\makeatletter

\renewcommand{\baselinestretch}{0.96}

\makeatletter
\def\ps@headings{%
\def\@oddhead{\mbox{}\scriptsize\rightmark \hfil }
\def\@evenhead{\scriptsize\thepage \hfil \leftmark\mbox{}}%
\def\@oddfoot{}%
\def\@evenfoot{}}
\makeatother
\usepackage{authblk}
\usepackage{epsfig}
\usepackage{epstopdf}
\usepackage{simplemargins}
\usepackage{amsfonts}
\usepackage{amsmath,bm}
\usepackage{graphicx}
\usepackage{url}
\usepackage{subfig}
\usepackage{booktabs}
\usepackage{multicol}
\usepackage[ruled,commentsnumbered]{algorithm2e}
\usepackage{algorithmic}
\usepackage{booktabs}
\usepackage{amssymb}
\usepackage{amsthm}

\usepackage{color}
\newcommand{\red}{\textcolor{black}}

\newtheorem{theorem}{Theorem}
\newtheorem{lemma}{Lemma}
\newtheorem{definition}{Definition}
\newtheorem{proposition}{Proposition}

\IEEEoverridecommandlockouts
\usepackage{cite}
\usepackage{algorithmic}
\usepackage{textcomp}
\def\BibTeX{{\rm B\kern-.05em{\sc i\kern-.025em b}\kern-.08em
		T\kern-.1667em\lower.7ex\hbox{E}\kern-.125emX}}
	
\begin{document}

\title{Joint Service Placement and Request Routing in Multi-cell Mobile Edge Computing Networks
\thanks{This publication was supported partly by the National Science Foundation under Grants CNS 1815676 and 1619129, and the U.S. Army Research Laboratory and the U.K. Ministry of Defence under Agreement Number W911NF-16-3-0001.}
}

\author[1]{Konstantinos Poularakis}
\author[2]{Jaime Llorca}
\author[2,3]{Antonia M. Tulino}
\author[4]{Ian Taylor}
\author[1]{Leandros Tassiulas\vspace{-0.5em}}
\affil[1]{Department of Electrical Engineering and Institute for Network Science, Yale University, USA} 
\affil[2]{Nokia Bell Labs, USA} 
\affil[3]{Department of Electrical Engineering and Information Technologies, University of Naples Federico II, Italy} 
\affil[4]{School of Computer Science and Informatics, Cardiff University, UK} 


\maketitle
\begin{abstract}
The proliferation of innovative mobile services such as augmented reality, 
\red{networked gaming, and autonomous driving}
has spurred a growing need for low-latency access to computing resources that cannot be met solely by existing 
\red{centralized cloud systems}. Mobile Edge Computing (MEC) is expected to be an effective solution to meet the demand for \red{low-latency services} by enabling the  execution of computing tasks at the network-periphery, in proximity to end-users.
\red{While a number of recent studies have addressed the problem of determining the execution of service tasks and the routing of user requests to corresponding edge servers, the focus has primarily been on
the efficient utilization of computing resources, neglecting the fact that non-trivial amounts of data need to be stored
to enable service execution, and that many emerging services exhibit asymmetric bandwidth requirements.}
To fill this gap, we study the joint optimization of service placement and request routing in MEC-enabled multi-cell networks with multidimensional (storage-computation-communication) constraints. We show that this problem generalizes several problems in literature and propose an algorithm that achieves close-to-optimal performance using randomized rounding. Evaluation results demonstrate that our approach can effectively 
\red{utilize the available resources to maximize  the number of requests served by low-latency edge cloud servers}.
\end{abstract}

\pagestyle{headings}
\pagenumbering{arabic}
\setcounter{page}{1}
\thispagestyle{empty}

\section{Introduction} \label{section:introduction}

\subsection{Motivation}

Emerging distributed cloud architectures, such as \emph{Fog} and \emph{Mobile Edge Computing (MEC)}, push substantial amounts of computing functionality to the edge of the network, in proximity to end-users, thereby allowing to bypass fundamental latency 
\red{limitations} of today's prominent centralized cloud systems~\cite{mec-survey}.
This trend is expected to continue unabated and play an important role in next-generation 5G networks for supporting both computation-intensive and latency-sensitive services~\cite{mec-5g}.

With MEC, services can be housed in wireless base stations (BSs) endowed with computing capabilities (or edge servers close to BSs) that can be used to accommodate service requests from users lying in their coverage regions. The computation capacity of BSs, however, is much more limited than that of centralized clouds, and may not suffice to satisfy all \red{user} 
requests. This naturally raises the question of \emph{where to execute} each service so as to better reap the benefits of available computation resources 
\red{to} serve as many 
requests as possible.

While there have been several interesting approaches to determine the execution (or offloading) of services in MEC, e.g.,~\cite{mec-dense} and~\cite{mec-iot}, to cite two of the most recent, an important aspect has been hitherto overlooked. Specifically, many services today require not only the allocation of computation resources, but also a \emph{non-trivial amount of data that needs to be pre-stored} (or pre-placed) at the BS. In an Augmented Reality (AR) service, for example, the placement of the object database and the visual recognition models is needed in order to run classification or object recognition before delivering the augmented information to the user~\cite{ar-uplink}. Yet, the storage capacity of BSs may be not large enough to support all \red{offered} services.

The above issue is further complicated by the communication requirements of the services. Many modern services require 
\red{uploading data from the user to be used as input for service execution, whose output must then be downloaded for consumption by the user.}
Such \emph{bidirectional communication} may be asymmetric in general, taking up different portions of \red{BSs'} uplink and downlink bandwidth capacities~\red{\cite{vr2018}}.

\begin{figure}[t]
	\begin{center}
		\includegraphics[scale=0.305]{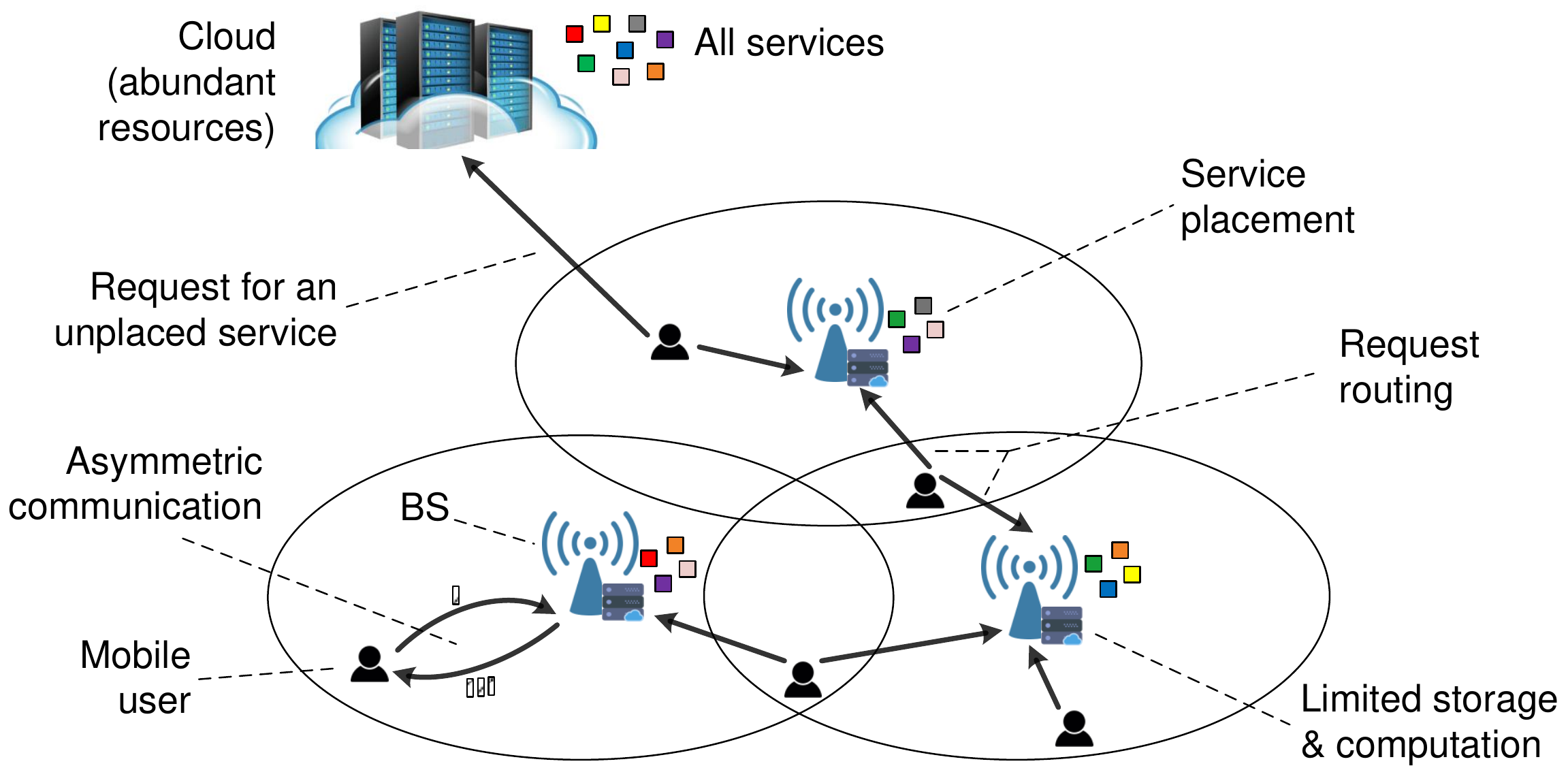}
		\caption{An example MEC system. Service placement and request routing are constrained by the storage, computation and bandwidth resources of BSs.}
		\label{fig:model}
	\end{center}
\vspace{-5mm}
\end{figure}

\red{In addition,} 
the density of BSs has been increasing and is expected to reach up to 50 BSs per km$^2$ in future 5G deployments~\cite{bs-density}. This \red{will} 
create a \emph{complex multi-cell environment} where users are concurrently in range of multiple BSs with overlapping coverage regions, and hence the operator can use multiple paths to route services to them. Figure \ref{fig:model} illustrates an example of such a system.

Evidently, \red{in this context}, MEC operators have a large repertoire of service placement and request routing alternatives for satisfying the user requests. In order to serve as many requests as possible from the BSs, the operator has to \emph{jointly} optimize these decisions \red{while simultaneously satisfying storage, computation, and communication constraints}.
Clearly, this is an important problem that differs substantially from previous related studies (e.g., see~\cite{mec-dense},~\cite{mec-iot} and the survey in~\cite{mec-survey}) that did not consider storage-constrained BSs and asymmetric communication requirements. While a few recent works~\cite{mec-storage},~\cite{mec-ting},~\cite{mec-cocaco} studied the impact of storage in MEC, they neither considered all the features of these systems discussed above nor provided \emph{optimal} or \emph{approximate} solutions for the joint service placement and request routing problem.

Given the above issues, the key open questions are:
\begin{itemize}
\item \emph{Which services to place in each BS to better utilize its available storage capacity?}
\item \emph{How to route user requests to BSs without overwhelming their available computation and (uplink/downlink) bandwidth capacities?}
\item \emph{How the above decisions can be optimized in a joint manner to offload the centralized cloud as much as possible?}
\end{itemize}

\subsection{Methodology and Contributions}
In this paper, we follow a systematic methodology in order to answer the above questions, summarized as follows.

\begin{enumerate}

\item We formulate the joint service placement and request routing problem (JSPRR) in multi-cell MEC networks aiming to minimize  the load of the centralized cloud. We consider practical features of these systems such as overlapping coverage regions of BSs and multidimensional (storage, computation, and communication) resource constraints.

\item We identify several placement and routing problems in literature that are special cases of JSPRR, gaining insights into the complexity of the original problem.

\item Using randomized rounding techniques~\cite{randomized-rounding}, we develop a bi-criteria algorithm that provably achieves approximation guarantees while violating the resource constraints in a bounded way. To the best of our knowledge, this is the first approximation algorithm for this problem.

\item We extend the results for dynamic scenarios where the user demand profiles change with time, and show how to adapt the solution accordingly.

\item We carry out evaluations to demonstrate the performance of the proposed algorithm. We show that, in many practical scenarios, our algorithm performs close-to-optimal and far better than \red{a state-of-the-art} 
method which neglects 
computation and bandwidth constraints.

\end{enumerate}

The rest of the paper is organized as follows. Section \ref{section:model} describes the system model and defines the JSPRR problem formally. We analyze the complexity of this problem and present algorithms with approximation guarantees in Section \ref{section:complexity} and \ref{section:approximation}, respectively.
Section \ref{section:extension} discusses extensions of our results and practical cases, while Section \ref{section:evaluation} presents our evaluation results. We conclude our work in Section \ref{section:conclusion}.


\section{Model and Problem Definition} \label{section:model}

We consider a MEC system consisting of a set $\mathcal{N}$ of $N$ BSs equipped with storage and computation capabilities and a set $\mathcal{U}$ of $U$ mobile users, subscribers of the MEC operator, as depicted in Figure \ref{fig:model}. The users can be arbitrarily distributed over the coverage regions of BSs, while the coverage regions may be overlapping in dense BS deployments. We denote by $\mathcal{N}_u \subseteq \mathcal{N}$ the \red{set of} BSs covering user $u$.

We consider multiple types of resources for BSs. First, each BS $n$ has \emph{storage capacity} $R_n$ (hard disk) that can be used to pre-store data associated with services. Second, BS $n$ has a CPU of \emph{computation capacity} (i.e., maximum frequency) $C_n$ that can be used to execute services in an on-demand manner. Third, BS $n$ has uplink (downlink) \emph{bandwidth capacity} $B^{\uparrow}_n$ ($B^{\downarrow}_n$) that can be used to upload (download) data from (to) mobile users requesting services.

The system offers a library $\mathcal{S}$ of $S$ services to the mobile users. Examples include video streaming, augmented reality \red{and networked} 
gaming. 
\red{S}ervices may have different requirements in terms of the amount of storage, CPU cycles, and uplink/downlink bandwidth. We denote by $r_s$ the storage space occupied by the data associated \red{with} 
service $s$. The notation $c_s$ indicates the required computation, while $b^{\uparrow}_s$ and $b^{\downarrow}_s$ indicate the uplink and downlink bandwidth required to satisfy a request for service $s$, respectively.

The system receives service requests from users in a stochastic manner. Without loss of generality, we assume that each user $u$ performs one request for a service denoted by $s_u$\footnote{If a user performs multiple requests, we can split it into multiple users.}. User requests can be predicted for a certain time period (e.g., a few hours) by using well-studied learning techniques (e.g., auto-regression analysis)~\cite{mec-storage}. Yet, the user demand can change after that period as the users may gain or lose interest in some services. We provide more details about this issue in Section \ref{section:extension}.

The request of user $u$ can be routed to a nearby BS in $\mathcal{N}_u$ provided that \red{service $s_u$} 
is locally \red{stored} 
and \red{the BS has} 
enough computation and bandwidth resources. If there is no such BS, we assume that the user can access the centralized cloud\red{,} which serves as a last resort for all users. Accessing the cloud, however, may cause high delay due to its long distance from \red{the} users, and therefore should be avoided.

The network operator needs to decide in which BSs to place the services and how to route user requests to them. To model these decisions, we introduce two sets of optimization variables: (i) $x_{ns} \in \{0,1\}$ which indicates whether service $s$ is placed in BS $n$ ($x_{ns}=1$) or not ($x_{ns}=0$), and (ii) $y_{nu} \in \{0,1\}$ which indicates whether the request of user $u$ is routed to BS $n$ ($y_{nu}=1$) or not ($y_{nu}=0$). Similarly, we denote by $y_{lu}$ the routing decision for the cloud. We refer by \emph{service placement} and \emph{request routing} policies to the respective vectors:
\begin{align}
\bm{x} &= ( x_{ns} \in \{0,1\}~: n \in \mathcal{N}, s \in \mathcal{S} ) \label{eq:x} \\
\bm{y} &= ( y_{nu} \in \{0,1\}~: n \in \mathcal{N}\cup\{l\}, u \in \mathcal{U} ) \label{eq:y}
\end{align}

The service placement and request routing policies need to satisfy several constraints. First, each user request needs to be routed to exactly one of the nearby BSs\red{,} or the  cloud:
\begin{equation}
\sum_{n\in \mathcal{N}_u \cup \{l\}} y_{nu} = 1, ~\forall u \in \mathcal{U} \label{eq:y1}
\end{equation}
Second, in order for the request of user $u$ to be routed to a BS $n$, the service $s_u$ needs to be placed in the latter:
\begin{equation}
y_{nu} \leq x_{ns_u}, ~\forall n \in \mathcal{N}, u \in \mathcal{U} \label{eq:xy}
\end{equation}
Third, the total amount of data of services placed in a BS must not exceed its storage capacity:
\begin{equation}
\sum_{s \in \mathcal{S}} x_{ns} r_s  \leq R_n, ~\forall n \in \mathcal{N}  \label{eq:storage}
\end{equation}
Fourth, the total computation load of user requests routed to BS $n$ must not exceed its computation capacity:
\begin{equation}
\sum_{u \in \mathcal{U}} y_{nu} c_{s_u} \leq C_n, ~\forall n \in \mathcal{N} \label{eq:computation}
\end{equation}
Fifth, the total bandwidth load of user requests routed to BS $n$ must not exceed its uplink and downlink bandwidth capacity:
\begin{align}
&\sum_{u \in \mathcal{U}}  y_{nu} b^{\uparrow}_{s_u} \leq B^{\uparrow}_n, ~\forall n \in \mathcal{N} \label{eq:bandwidth-up} \\
&\sum_{u \in \mathcal{U}}  y_{nu} b^{\downarrow}_{s_u} \leq B^{\downarrow}_n, ~\forall n \in \mathcal{N} \label{eq:bandwidth-down}
\end{align}

The goal of the network operator is to find the joint service placement and request routing policy that maximizes the number of requests served by the BSs\red{,} or\red{,} equivalently\red{,} minimizes the load of the cloud:
\begin{eqnarray}
\min _{ \bm{x}, \bm{y} }  &  \sum_{u \in \mathcal{U}} y_{lu}   \\
\text{s.t.} & \text{constraints:}~(\ref{eq:x}) ~-~ (\ref{eq:bandwidth-down})\nonumber
\end{eqnarray}
We refer by \emph{JSPRR} to the above problem. This is an integer optimization problem and such problems are typically challenging to solve. In the next two sections, we analyze the complexity of this problem and propose approximation algorithms. 

\section{Complexity Analysis} \label{section:complexity}

The JSPRR problem is \emph{NP-Hard} since it generalizes the knapsack problem~\cite{knapsack} by comprising multiple packing constraints (inequalities (\ref{eq:storage})-(\ref{eq:bandwidth-down})). The problem remains challenging even when simplified by the assumption of \emph{homogeneous (unit-sized) service requirements}, i.e., $r_s=c_s=b^{\uparrow}_s=b^{\downarrow}_s=1$ $\forall s$. This simplified problem includes as special cases several \emph{well-studied placement and routing problems} in literature, which shows the universality of our model. Subsequently, we describe several such special cases. We begin with the (rather trivial) special case of non-overlapping BS coverage regions before diving into more challenging special cases.

\subsection{Special case 1: Non-overlapping BS coverage regions} \label{section:complexity1}

In the first special case, we make the simplifying assumption (in addition to the homogeneity of service requirements) that the coverage regions of the BSs do not overlap with each other. This particularly applies to sparse BS deployments where the BSs are located far away one from the other. It follows that the JSPRR problem can be \emph{decomposed into $N$ independent subproblems}, one per BS $n$.
\red{The objective of subproblem $n$ is to maximize the number of requests served by BS $n$.}

It is not difficult to show that there is always an optimal solution to subproblem $n$ that places in BS $n$ the $R_n$ most \red{locally} popular services, i.e., the services requested by most users inside the coverage region of BS $n$. Then, 
BS $n$ will admit as many request\red{s} for the placed services as its computation and bandwidth capacities $C_n$, $B^{\uparrow}_n$ and $B^{\downarrow}_n$ can handle, i.e., $\min\{C_n, B^{\uparrow}_n, B^{\downarrow}_n\}$ requests at most. Indeed, consider a solution that places in BS $n$ a service $s_1$ requested by fewer users inside the respective coverage region \red{than} 
another service $s_2$. Then, one could swap the two services in the placement solution and route the same number of 
requests to BS $n$ without changing the objective function value. Therefore, the JSPRR problem is trivial to solve in this special case.

\subsection{Special case 2: Data placement/caching problem}

In the second special case, we allow the coverage regions of BSs to \red{overlap}, but we make the simplifying assumption that the computation and bandwidth resources are non-congestible, i.e., they always suffice to route all user requests to BSs. In other words, we assume that the capacities $C_n$, $B^{\uparrow}_n$ and $B^{\downarrow}_n$ are greater \red{than} or equal to the demand of users, so \red{that} we can remove 
constraints 
(\ref{eq:computation})-(\ref{eq:bandwidth-down}) from the problem formulation without affecting the optimal solution.

Without the computation and bandwidth constraints, the placement and routing problem becomes much simpler. For a given service placement solution $\bm{x}$, finding the optimal request routing policy $\bm{y}$ is straightforward; simply route each user request to a nearby BS having stored the requested service, if any, otherwise to the cloud. This special case has been extensively studied in literature under the title \emph{`data placement'}~\cite{data-placement} or \emph{`caching' problem}~\cite{femtocaching}. This problem asks to place data items (services) to caches (BSs) with the objective of maximizing the total number of requests served by the caches.

While the data placement problem is NP-Hard, several approximation algorithms are known in literature. The main method used to derive such approximations is based on showing the submodularity property of the \red{optimization problem}. 
That is\red{,} to show that the marginal value of the 
objective function never increases as more data items are placed in the caches. Having shown the submodularity property, several `classic' algorithms can be applied, with the most known being greedy, local search\red{,} and pipage rounding~\cite{femtocaching}. Among the three algorithms, the greedy is the simplest and fastest\red{,} and, hence, the most practical. 

\subsection{Special case 3: Middlebox placement problem}

In the third special case, we allow the coverage regions of BSs to overlap and the computation and bandwidth resources to be congestible\red{,} but we make the simplifying assumption that the storage capacities are unit-sized ($\red{R_n}=1$ $\forall n$). That is, we assume that only one service can be stored per BS.

Under this special case, the JSPRR problem can be reduced to the \emph{`middlebox placement' problem}~\cite{match},~\cite{incremental}. While there exist many different variants of the middlebox placement problem in literature, typically, this problem asks to pick $l$ out of $m$ nodes in a network to deploy middleboxes. The goal is to maximize the total number of source-destination flows (out of $f$ flows) that can be routed through network paths containing at least one middlebox, subject to a constraint that limits the number of flows per middlebox. 

Although the reduction is not so straightforward, the main idea is to create: (i) a distinct node for each pair of a BS and a service  ($m = N S$) and (ii) a distinct flow for each user ($f=U$). Each flow can be routed through any node whose BS-service pair satisfies that the BS covers the respective user and the service is the requested by the user. The question is which $l=N$ out of the $m=NS$ nodes to pick to deploy middleboxes, with an additional constraint per BS that only $1$ our of the $S$ nodes corresponding to that BS can be picked. The picked node will determine which of the $S$ services is placed at that BS.

Recent works have shown that the maximum flow objective of the middlebox problem is a submodular function~\cite{match},~\cite{incremental}. Therefore, this problem can be solved by using the same approximation algorithms mentioned in special case 2.


\subsection{General case: Non-submodular}
Although it would be tempting to conjecture that our JSPRR problem is submodular in its general form (with overlapping coverage regions, congestible bandwidth and computation and large storage capacities), we can construct counter-examples where this property does not hold. First, we introduce the definition of submodular functions.

\begin{definition}
Given a finite set of elements $\mathcal{G}$ (ground set), a function $f: 2^{\mathcal{G}} \rightarrow \mathbb{R}$ is submodular if for any sets $\mathcal{A}\subseteq \mathcal{B} \subseteq \mathcal{G}$ and every element $g \notin \mathcal{B}$, it holds that:
\begin{equation}
f(\mathcal{A} \cup \{g\}) - f(\mathcal{A}) \geq f(\mathcal{B} \cup \{g\}) - f(\mathcal{B})
\end{equation}
\end{definition}

Next, we introduce the element $e_{ns}$ to denote the placement of service $s$ in BS $n$. The ground set is given by $\{e_{11},\dots,e_{NS}\}$. Every possible service placement policy can be expressed by a subset $\mathcal{E}\subseteq \mathcal{G}$ of elements, where the elements included in $\mathcal{E}$ correspond to the service placement.
Given a service placement $\mathcal{E}$, we denote by $f(\mathcal{E})$ the maximum number of user requests that can be satisfied by the BSs.


We will construct a counter-example where the function $f(\mathcal{E})$ is not submodular. Specifically, we consider a system of $N=2$ BSs and $U=2$ users located in the intersection of the two coverage regions. The users request two different services denoted by $s_1$ and $s_2$. We set the computation capacities to $C_1=C_2=1$ (i.e., at most one service request can be satisfied by each BS), while the storage and bandwidth capacities are abundant. The two placement sets we consider are $\mathcal{A} = \{e_{11}\}$ and $\mathcal{B} =  \{e_{11}, e_{21}\}$, where $\mathcal{A} \subseteq \mathcal{B}$. We note that $f(\mathcal{A})=f(\mathcal{B})=1$ since in both cases only one of the two services is stored ($s_1$), and hence only one of the two requests can be served. Besides, $f(\mathcal{A}\cup\{e_{12}\})=1$ since the computation constraint prevents the BS $1$ from serving both user requests. However, $f(\mathcal{B}\cup\{e_{12}\})=2$ since now each BS can serve one user request. Therefore, the marginal performance is larger for the set $\mathcal{B}$ than the $\mathcal{A}$, which means that $f$ is not submodular.

We note that similar counter-examples can be identified when the bandwidth (instead of computation) capacity is congestible, provided that storage capacity is greater than $1$.

\begin{figure}[t]
	\begin{center}
		\includegraphics[scale=0.45]{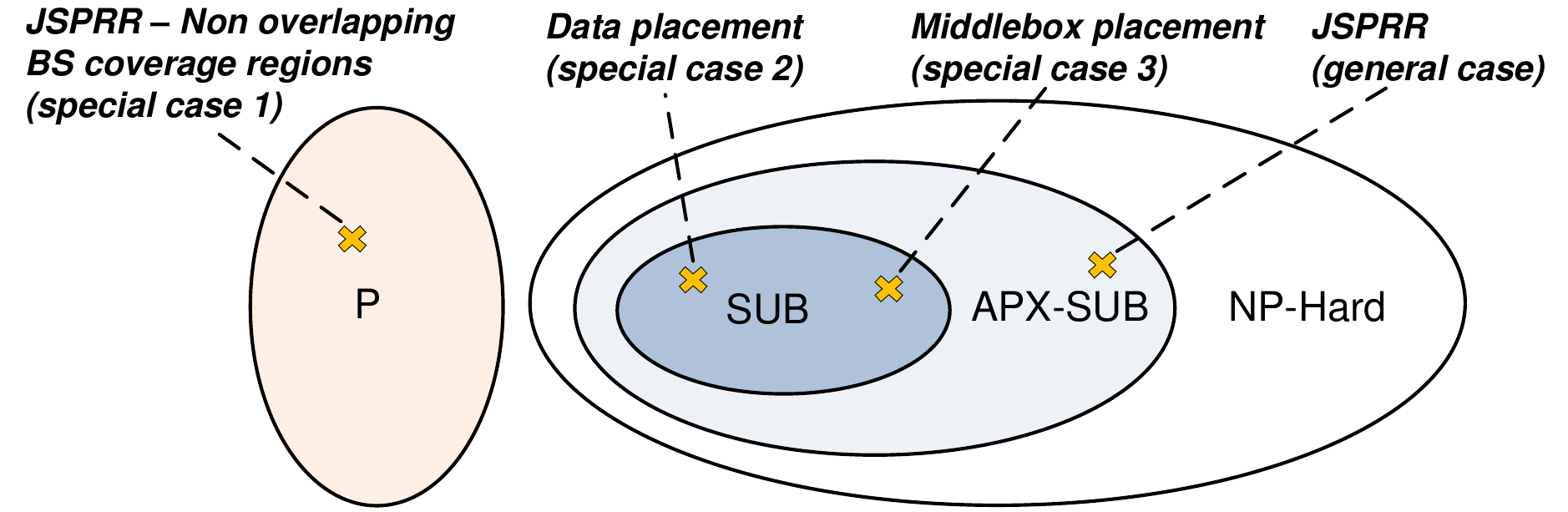}
		\caption{Complexity of special cases of JSPRR: Polynomial-time solvable (P), Submodular (SUB) and Approximately submodular (APX-SUB) classes.}
		\label{fig:complexity}
	\end{center}
\vspace{-4mm}
\end{figure}

\subsection{General case: Approximately-submodular}

Although our JSPRR problem does not fall into the class of submodular problems, we can show that it belongs to the wider class of \emph{approximately submodular} problems~\cite{e-submodular}. The complexity of JSPRR for the general and special cases is illustrated in Figure \ref{fig:complexity}.

\begin{definition}
A function $f: 2^{\mathcal{G}} \rightarrow \mathbb{R}$ is $\delta$-approximately submodular if there exists a submodular function $F: 2^{\mathcal{G}} \rightarrow \mathbb{R}$ such that for any $\mathcal{E}\subseteq \mathcal{G}$:
\begin{equation}
(1-\delta) F( \mathcal{E} ) \leq  f(\mathcal{E}) \leq (1+\delta)F(\mathcal{E}) \label{eq:delta}
\end{equation}
\end{definition}

We define by $F(\mathcal{E})$ the maximum number of user requests that can be satisfied by the BSs given the service placement set $\mathcal{E}$ in the special case that the bandwidth and computation resources are non-congestible (special case 2). Since there are fewer constraints in this special case than in the general case, it holds that $ f(\mathcal{E}) \leq F(\mathcal{E}) $. Therefore, for any $\delta\in[0,1]$, we have $ f(\mathcal{E}) \leq (1+\delta) F(\mathcal{E}) $. What remains to find is a $\delta$ value that satisfies the first inequality in (\ref{eq:delta}), i.e., $(1-\delta) F(\mathcal{E}) \leq f(\mathcal{E})$.

We note that when computing the value of $F(\mathcal{E})$, the BS $n$ is allowed to satisfy all the requests for stored services generated by users in its coverage region. We denote by $\Phi_n$ the number of these requests. In case that it happens $\Phi_n\leq C_n$,  $\Phi_n\leq B^{\uparrow}_n$ and $\Phi_n\leq B^{\downarrow}_n$ $\forall n$, then the computation and bandwidth resources are non-congestible and we have $f(\mathcal{E})=F(\mathcal{E})$. In the other case that, for some $n$, it happens $\Phi_n > C_n$ or $\Phi_n > B^{\uparrow}_n$ or $\Phi_n > B^{\downarrow}_n$, then the BS $n$ can process up to $\Phi_n/C_n$ times more requests, compared to $f(\mathcal{E})$. Similarly, the BS $n$ can receive (deliver) data from (to) up to $\Phi_n/B^{\uparrow}_n$ ($\Phi_n/B^{\downarrow}_n$) times more users. Therefore, the total number of satisfied requests is upper bounded by:
\begin{align}
F(\mathcal{E}) &\leq \max_{n\in\mathcal{N}}\{ \frac{\Phi_n}{C_n}, ~ \frac{\Phi_n}{B^{\uparrow}_n}, ~ \frac{\Phi_n}{B^{\downarrow}_n},~ 1 \} f(\mathcal{E})
\end{align}
where the value $1$ inside the $\max$ operator ensures that $F(\mathcal{E})$ will never be lower than $f(\mathcal{E})$. We thus can ensure that $(1-\delta) F(\mathcal{E}) \leq f(\mathcal{E})$ by picking:
\begin{align}
\delta =  1 - \frac{ 1 } { \max_{n\in\mathcal{N}}\{ \frac{\Phi_n}{C_n}, ~ \frac{\Phi_n}{B^{\uparrow}_n}, ~ \frac{\Phi_n}{B^{\downarrow}_n}, ~1 \} }
\end{align}

The problem of maximizing a $\delta$-approximately submodular function has been studied in the past~\cite{e-submodular}. Based on the results in~\cite{e-submodular}, we can use a simple greedy algorithm to achieve the approximation ratio described in the following proposition.

\begin{proposition}
The Greedy algorithm returns a solution set $\mathcal{E}^*$ such that:
\begin{align}
f(\mathcal{E}^*) \geq \frac{1}{2} \Big(\frac{1-\delta}{1+\delta}\Big)\frac{1}{1+\frac{\sum_{n\in\mathcal{N}}R_n \delta}{1-\delta}} \max_{\mathcal{E}} f(\mathcal{E})
\end{align}
\end{proposition}

Consider for example the case that the demand exceeds the available resources by up to $50\%$, i.e., there exist\red{s a BS} $n$ for which $\Phi_n=1.5C_n$ or $\Phi_n=1.5B^{\uparrow}_n$ or $\Phi_n=1.5B^{\downarrow}_n$. Then, 
$\delta = 1/3$, and the approximation factor \red{becomes:} 
\begin{align}
f(\mathcal{E}^*) \geq \frac{1}{4} \frac{1}{1+\frac{\sum_{n\in\mathcal{N}}R_n}{2}} \max_{\mathcal{E}} f(\mathcal{E})
\end{align}
The above approximation ratio worsens as the network becomes more congested ($\delta$ increases) and the storage capacities increase ($R_n$). This suggests that the JSPRR problem is substantially harder than the placement and routing problems described above, and thus a method of different philosophy is needed to find a tight approximation ratio. In the next section, we 
present such a method that goes beyond submodularity and achieves a solution with better approximation guarantees.

\section{Approximation Algorithm} \label{section:approximation}

In this section, we present one of the main contributions of this work; a novel approximation algorithm for the JSPRR problem. Our algorithm leverages a Randomized Rounding technique, from which it takes its name. The algorithm is described in detail below and summarized in Algorithm \ref{alg:rr}.

\red{The} Randomized Rounding algorithm starts by solving the linear relaxation of the JSPRR problem (line 1). That is, it relaxes the variables \{$x_{ns}$\} and \{$y_{nu}$\} to be fractional, rather than integer. The Linear Relaxation of JSPRR problem, \emph{LR-JSPRR} for short, can be expressed as follows:
\begin{eqnarray}
\min _{ \bm{x}, \bm{y} }  &  \sum_{u \in \mathcal{U}} y_{lu}   \\
\text{s.t.} & \text{constraints:}~(\ref{eq:y1}) ~-~ (\ref{eq:bandwidth-down})\nonumber \\
     & x_{ns} \in [0,1], ~\forall n \in \mathcal{N}, s \in \mathcal{S} \label{eq:xrelaxed} \\
     & y_{nu} \in [0,1], ~\forall n \in \mathcal{N}\cup\{l\}, u \in \mathcal{U} \label{eq:yrelaxed}
\end{eqnarray}
where we have replaced equations (\ref{eq:x})-(\ref{eq:y}) with (\ref{eq:xrelaxed})-(\ref{eq:yrelaxed}). Since the objective and the constraints of the above problem are linear, it can be optimally solved in polynomial time using a linear program solver. We denote by $\{x^{\dag}_{ns}\}$ and $\{y^{\dag}_{nu}\}$ the optimal solution values. The next step is to round these values \red{to obtain an integer solution,} 
denoted by $\{\widehat{x}_{ns}\}$ and $\{\widehat{y}_{nu}\}$. For each pair of node $n$ and service $s$\red{,} the algorithm \red{rounds} 
variable $\widehat{x}_{ns}$ to 1 with probability $x^{\dag}_{ns}$ (lines 2-3). Each rounding decision \red{is taken independently from each other.} 

Finally, the algorithm uses the rounded placement variable\red{s} 
$\{\widehat{x}_{ns}\}$ to decide the rounding of the routing variables (lines 4-9). For each user $u$, it defines the set of nearby BSs that have stored the requested service $s_u$ by $\mathcal{N}'_u = ( n \in \mathcal{N}_u ~:~ \widehat{x}_{ns_u} > 0 ) $ and uses this to distinguish between two cases:
(i) if user $u$ cannot find service $s_u$ in any of the nearby BSs ($\mathcal{N}'_u=\emptyset$), then the user is served by the  cloud (lines 6-7),
(ii) otherwise,  the user is randomly routed to one of the BSs in $\mathcal{N}'_u$ or the  cloud (lines 8-9). The \red{routing} probabilities 
depend on the fractional values $\{x^{\dag}_{ns}\}$ and $\{y^{\dag}_{nu}\}$. Higher probability is given to BSs with larger $y^{\dag}_{nu}$ values.

\begin{algorithm}[t]
\BlankLine
\nl Solve the linear relaxation of JSPRR problem to obtain $(\bm{x}^{\dag},\bm{y}^{\dag})$ optimal solution\\%
\nl \For{$n\in \mathcal{N}$, $s \in \mathcal{S}$}{
\nl Set $\widehat{x}_{ns} = 1$ with probability $x^{\dag}_{ns}$\\}
\nl \For{$u\in \mathcal{U}$}{
\nl Define $\mathcal{N}'_u = ( n \in \mathcal{N}_u ~:~ \widehat{x}_{ns_u} > 0 )$ and  \\
\nl \eIf{$ \mathcal{N}'_u = \emptyset $}{
\nl set $\widehat{y}_{lu} = 1$ and $\widehat{y}_{nu} = 0$ $\forall n \in \mathcal{N}$}
  {
  \nl set $\widehat{y}_{nu} = 1$, $n\in\mathcal{N}'_u$, with probability $\frac{ y^{\dag}_{nu} } { x^{\dag}_{ns_u} }$, or \\
  \nl set $\widehat{y}_{lu} = 1$ with probability $\bigg[\frac{ y^{\dag}_{lu} - \prod_{n\in\mathcal{N}'_u} ( 1-x^{\dag}_{ns_u} ) }{1 - \prod_{n \in \mathcal{N}'_u}(1- x^{\dag}_{ns_u}) } \bigg]_+$
   }
 }
\nl Output $\widehat{\bm{x}},\widehat{\bm{y}}$\\%
\caption{Randomized Rounding algorithm} \label{alg:rr}
\end{algorithm}

Subsequently, we provide guarantees on the quality of the solution returned by \red{the} Randomized Rounding algorithm. We begin with the following lemma.

\begin{lemma}
\red{The} Randomized Rounding algorithm routes all user requests with high probability. 
\end{lemma}
\begin{proof}
For a given user $u$, there are two cases when rounding the fractional variable $y^{\dag}_{nu}$ to $\widehat{y}_{nu}$: (i) there is no nearby BS having stored the requested service ($\mathcal{N}'_u = \emptyset$) and (ii) there is at least one such BS ($\mathcal{N}'_u \neq \emptyset$). The probability that the request of user $u$ is routed to the cloud is given by:

\begin{align}
\Pr[\widehat{y}_{lu}=1] & = \Pr\Big[\widehat{y}_{lu}=1 ~|~ \mathcal{N}'_u =\emptyset \Big] \Pr\Big[\mathcal{N}'_u =\emptyset \Big] \nonumber \\
                        & + \Pr\Big[\widehat{y}_{lu}=1 ~|~ \mathcal{N}'_u \neq \emptyset \Big] \Pr\Big[\mathcal{N}'_u \neq \emptyset \Big] \nonumber \\
                        & = 1 \prod_{n \in \mathcal{N}'_u}(1-x^{\dag}_{ns_u})  \nonumber \\
                        & + \frac{ y^{\dag}_{lu} - \prod_{n\in\mathcal{N}'_u} ( 1-x^{\dag}_{ns_u} ) }{1 - \prod_{n \in \mathcal{N}'_u}(1- x^{\dag}_{ns_u})}(1 - \prod_{n \in \mathcal{N}'_u}(1- x^{\dag}_{ns_u})) \nonumber \\
                        & = y^{\dag}_{lu} \label{eq:probyb}
\end{align}
The first equation is by the definition of conditional probability. The second equation is because the request of user $u$ will be routed with probability $1$ to the cloud if $\mathcal{N}'_u=\emptyset$, and the $\{x^{\dag}_{ns}\}$ variables are rounded independently \red{of} one another (hence, $\Pr[\mathcal{N}'_u =\emptyset] = \prod_{n \in \mathcal{N}'_u}(1-x^{\dag}_{ns_u})$). To simplify the analysis, we have assumed that $y^{\dag}_{lu} \geq \prod_{n\in\mathcal{N}'_u} ( 1-x^{\dag}_{ns_u} )$ and, thus, removed the $[.]_+$ operator from the respective probability. This condition is expected to hold in practice as the BS deployment becomes more and more dense (larger $\mathcal{N}'_u$ sets) and the BS storage capacities increase (larger $x^{\dag}_{ns_u}$ values).

Similarly, the probability that user $u$ is routed to BS $n$ is:
\begin{align}
\Pr[\widehat{y}_{nu}=1] &= \Pr\Big[\widehat{y}_{nu}=1 ~|~ \widehat{x}_{ns_u}=1 \Big] \Pr\Big[ \widehat{x}_{ns_u}=1 \Big]  \nonumber \\
& + \Pr\Big[\widehat{y}_{nu}=1 ~|~ \widehat{x}_{ns_u}=0 \Big] \Pr\Big[ \widehat{x}_{ns_u}=0 \Big]  \nonumber \\
& = \frac{ y^{\dag}_{nu} } { x^{\dag}_{ns_u}  }  x^{\dag}_{ns_u}  = y^{\dag}_{nu} \label{eq:probyn}
\end{align}

The sum of probabilities of routing the request of user $u$ to the  cloud or the BSs is:
\begin{align}
\sum_{ n \in \mathcal{N}_u \cup \{l\} } \Pr[\widehat{y}_{nu}=1] =  \sum_{n \in \mathcal{N}_u \cup \{l\} } y^{\dag}_{nu} = 1 \label{eq:almost1}
\end{align}
where the last equation holds due to (\ref{eq:y1}). The above is very close to the probability of routing the request of user $u$ except for an additive gap that converges to zero  as the number of  BSs covering user $u$ increases.
\end{proof}

By construction, Randomized Rounding 
routes requests only to BSs that have stored the respective service (in $\mathcal{N}'_u$ set in line 5) or \red{to} 
the cloud. Therefore, constraint (\ref{eq:xy}) is also satisfied. Next, we study whether the remaining constraints in (\ref{eq:storage}), (\ref{eq:computation}), (\ref{eq:bandwidth-up}) and (\ref{eq:bandwidth-down})  are satisfied.

\begin{lemma}
The solution returned by \red{the} Randomized Rounding algorithm satisfies in expectation the storage, computation and bandwidth capacity constraints in (\ref{eq:storage}), (\ref{eq:computation}), (\ref{eq:bandwidth-up})\red{,} and (\ref{eq:bandwidth-down}).
\end{lemma}
\begin{proof}
We begin with the storage capacity constraint. The expected amount of data placed in BS $n$ is given by:
\begin{equation}
\mathbb{E}[\sum_{ s \in \mathcal{S} } \widehat{x}_{ns} r_s ] = \sum_{ s \in \mathcal{S} } \Pr[\widehat{x}_{ns}=1] r_s = \sum_{ s \in \mathcal{S} }x^{\dag}_{ns} r_s =  R_n \label{eq:exR}
\end{equation}
where the second equation is because the \red{$\{\widehat{x}_{ns}\}$} variables are binary\red{,} with success probabilities the fractional values \red{$\{x^{\dag}_{ns}\}$}. The last equation is \red{due to} constraint (\ref{eq:storage}) and the fact that it would be wasteful to not use all the storage space.

Next, we consider the computation capacity constraint. The expected computation load of BS $n$ is given by:

{\small
\begin{align}
\mathbb{E}[\sum_{u \in \mathcal{U}} \widehat{y}_{nu} c_{s_u}] =  \sum_{u \in \mathcal{U}} \Pr[\widehat{y}_{nu}=1] c_{s_u} =  \sum_{u \in \mathcal{U} } y^{\dag}_{nu} c_{s_u} \leq C_n \label{eq:exC}
\end{align}
}
where the second equation holds due to equation (\ref{eq:probyn}). The inequality is by constraint (\ref{eq:computation}). Similar inequalities can be shown for the uplink/downlink bandwidth constraints:

{\small
\begin{align}
\mathbb{E}[\sum_{u \in \mathcal{U}} \widehat{y}_{nu} b^{\uparrow}_{s_u}] =  \sum_{u \in \mathcal{U}} \Pr[\widehat{y}_{nu}=1]  b^{\uparrow}_{s_u} =  \sum_{u \in \mathcal{U} } y^{\dag}_{nu}  b^{\uparrow}_{s_u} \leq B^{\uparrow}_n \label{eq:exBup} \\
\mathbb{E}[\sum_{u \in \mathcal{U}} \widehat{y}_{nu} b^{\downarrow}_{s_u}] =  \sum_{u \in \mathcal{U}} \Pr[\widehat{y}_{nu}=1]  b^{\downarrow}_{s_u} =  \sum_{u \in \mathcal{U} } y^{\dag}_{nu}  b^{\downarrow}_{s_u} \leq B^{\downarrow}_n \label{eq:exBdown}
\end{align}
}
where we have used equations (\ref{eq:probyn}), (\ref{eq:bandwidth-up})\red{,} and (\ref{eq:bandwidth-down}).
\end{proof}

A similar result holds for the objective function value.
\begin{lemma}
The objective value returned by \red{the} Randomized Rounding algorithm is in expectation equal to \red{that} 
of the optimal fractional solution.
\end{lemma}
\begin{proof}
The expected number of user requests routed to the cloud by Randomized Rounding 
is given by:
\begin{align}
& \mathbb{E}[\sum_{u \in \mathcal{U}} \widehat{y}_{lu}] = \sum_{u \in \mathcal{U}} \Pr[\widehat{y}_{lu}=1] =  \sum_{u \in \mathcal{U}} y^{\dag}_{lu}
\end{align}
where the second equation holds due to equation (\ref{eq:probyb}).
\end{proof}

The above lemmas have shown that \red{the} Randomized Rounding algorithm satisfies \emph{in expectation} all the constraints and achieves the optimal \red{objective function} 
value. However, \emph{in practice}, the constraints may be violated. Therefore, it is important to bound the factor by which this happens.

\begin{theorem} \label{theorem:1}
The amount of data placed by \red{the} Randomized Rounding algorithm in BS $n$ will not exceed its storage capacity by a factor \red{larger than} $\frac{ 3 \ln (S)}{R_n} +4 $ with high probability.
\end{theorem}
\begin{proof}
For a given BS $n$, the products $ \widehat{x}_{ns} r_s$ $\forall s\in\mathcal{S}$ are independent random variables with expected total value $\mathbb{E}[\sum_{s\in\mathcal{S}}\widehat{x}_{ns} r_s] =  R_n$ (cf. equation (\ref{eq:exR})). Moreover, by appropriately normalizing the $r_s$ and $R_n$ values, we can ensure that the $\widehat{x}_{ns} r_s$ variables take values within $[0,1]$. Therefore, we can apply the Chernoff Bound theorem~\cite{chernoff} to show that for any $\epsilon>0$:
\begin{equation}
\Pr [ \sum_{ s \in \mathcal{S} } \widehat{x}_{ns} r_s \geq (1+\epsilon) R_n ]  \leq \exp^{\frac{-\epsilon^2 R_n}{2+\epsilon}}
\end{equation}
Next, we find an $\epsilon$ value for which the probability upper bound above becomes very small. Specifically, we require that:
\begin{equation}
\exp^{\frac{-\epsilon^2 R_n}{2+\epsilon}} \leq \frac{1}{S^3} \label{eq:boundR}
\end{equation}
which means that the probability bound goes quickly (at a cubic rate) to zero as the number of services increases. In order for this to be true, the $\epsilon$ value must satisfy:
\begin{equation}
\epsilon \geq \frac{3\ln(S)}{2R_n}  +\sqrt{   \frac{9\ln^2(S)}{4R^2_n}  +  \frac{6\ln(S)}{R_n}     }
\end{equation}
The above condition holds if we pick:
\begin{equation}
\epsilon= \frac{3\ln(S)}{R_n} +3
\end{equation}
since, in practice, $R_n\geq \ln(S)$. Finally, we upper bound the probability that \emph{any} of the BS storage capacities is violated:
\begin{align}
&\Pr [ \bigcup_{n\in\mathcal{N}} \sum_{ s \in \mathcal{S} } \widehat{x}_{ns} r_s \geq (1+\epsilon) R_n       ]  \nonumber\\
&\leq \sum_{n\in\mathcal{N}} \Pr [  \sum_{ s \in \mathcal{S} } \widehat{x}_{ns} r_s \geq (1+\epsilon) R_n    ]  \nonumber\\
&\leq N \frac{1}{S^3} \leq \frac{1}{S^2}
\end{align}
where the first inequality is due to the Union Bound theorem. The second inequality is due to inequality (\ref{eq:boundR}) and because the number of BSs is $N$. The last inequality is because, in practice, the service library size is larger than the number of BSs ($S>N$). Therefore, with high probability, the storage capacity of any BS $n$ will not be exceeded by more than a factor of $1+\epsilon = \frac{ 3 \ln (S)}{R_n} +4 $.
\end{proof}

\begin{theorem} \label{theorem:2}
The computation load of BS $n$ returned by \red{the} Randomized Rounding algorithm will not exceed its capacity by a factor of $\frac{3\ln(S)}{\lambda^{\dag}}+4$  with high probability, where $\lambda^{\dag}$ is the minimum computation load among BSs in the optimal fractional solution. 
\end{theorem}
\begin{proof}
The proof is similar to Theorem \ref{theorem:1}. For a given BS $n$, the variables $\widehat{y}_{nu} c_{s_u}$ $\forall u\in\mathcal{U}$ are independent with expected total value $\mathbb{E}[\sum_{u\in\mathcal{U}}\widehat{y}_{nu}c_{s_u}] =  \sum_{u \in \mathcal{U} } y^{\dag}_{nu}c_{s_u} $ (cf. inequality (\ref{eq:exC})). Moreover, they can be normalized to take values within $[0,1]$. Therefore, we can apply the Chernoff Bound theorem:

{\small
\begin{equation}
\Pr [ \sum_{ u \in \mathcal{U} } \widehat{y}_{nu} c_{s_u} \geq (1+\epsilon)  \sum_{u \in \mathcal{U} } y^{\dag}_{nu} c_{s_u}    ]  \leq  \exp^{\frac{-\epsilon^2 \sum_{u \in \mathcal{U} } y^{\dag}_{nu} c_{s_u} }{2+\epsilon}} \label{eq:boundC}
\end{equation}
}
Unlike storage, however, the expected computation load may not be equal to the capacity, i.e., $\sum_{u \in \mathcal{U} } y^{\dag}_{nu} c_{s_u} \neq C_n $. Therefore, we cannot replace it in the above inequality. To overcome this obstacle, we use the fact that $\sum_{u \in \mathcal{U} } y^{\dag}_{nu} c_{s_u} \leq C_n $ (by constraint (\ref{eq:computation})) and $\lambda^{\dag} \leq \sum_{u \in \mathcal{U} } y^{\dag}_{nu} c_{s_u}$ (by definition of $\lambda^{\dag}$) to show the following two inequalities:

{\footnotesize
\begin{align}
\Pr [ \sum_{ u \in \mathcal{U} } \widehat{y}_{nu} c_{s_u} \geq (1+\epsilon)  C_n     ]  &\leq  \Pr [ \sum_{ u \in \mathcal{U} } \widehat{y}_{nu} c_{s_u} \geq (1+\epsilon) \sum_{u \in \mathcal{U} }  y^{\dag}_{nu} c_{s_u}     ]  \label{eq:boundC1} \\
\exp^{\frac{-\epsilon^2 \sum_{u \in \mathcal{U} } y^{\dag}_{nu} c_{s_u} }{2+\epsilon}}  &\leq  \exp^{\frac{-\epsilon^2   \lambda^{\dag}   }{2+\epsilon}} \label{eq:boundC2}
\end{align}
}
By combining inequalities (\ref{eq:boundC}), (\ref{eq:boundC1})\red{,} and (\ref{eq:boundC2}), we obtain:
\begin{equation}
\Pr [ \sum_{ u \in \mathcal{U} } \widehat{y}_{nu} c_{s_u} \geq (1+\epsilon)  C_n     ]  \leq  \exp^{\frac{-\epsilon^2  \lambda^{\dag}   }{2+\epsilon}} \label{eq:boundC3}
\end{equation}
To complete the proof, we will find an $\epsilon$ value for which the probability upper bound above becomes very small, i.e., at most $1/S^3$. Similarly to Theorem \ref{theorem:1}, we can set $\epsilon= \frac{3\ln(S)}{ \lambda^{\dag} } +3$. Then, we can upper bound the probability that \emph{any} of the computation capacities is violated by:
\begin{align}
&\Pr [ \bigcup_{n\in\mathcal{N}} \sum_{ u \in \mathcal{U} } \widehat{y}_{nu} c_{s_u} \geq (1+\epsilon)   C_n     ]  \nonumber\\
&\leq \sum_{n\in\mathcal{N}} \Pr [  \sum_{ u \in \mathcal{U} } \widehat{y}_{nu} c_{s_u} \geq (1+\epsilon) C_n    ]  \nonumber\\
&\leq N \frac{1}{S^3} \leq \frac{1}{S^2}
\end{align}
This means that, with high probability, the computation capacity of any BS $n$ will not be exceeded by more than a factor of $1+\epsilon = \frac{ 3 \ln (S)}{\lambda^{\dag}} +4 $.
\end{proof}

Using similar arguments, the following two theorems can be proved for the uplink and downlink bandwidth capacities.

\begin{theorem}\label{theorem:3}
The uplink bandwidth load of BS $n$ returned by \red{the} Randomized Rounding algorithm will not exceed its capacity by a factor of $\frac{3\ln(S)}{\mu^{\dag}}+4$ with high probability, where $\mu^{\dag}$ is the minimum uplink bandwidth load among BSs in the optimal fractional solution. 
\end{theorem}

\begin{theorem}\label{theorem:4}
The downlink bandwidth load of BS $n$ returned by the Randomized Rounding algorithm will not exceed its capacity by a factor of $\frac{3\ln(S)}{\nu^{\dag}}+4$ with high probability, where $\nu^{\dag}$ is the minimum downlink bandwidth load among BSs in the optimal fractional solution. 
\end{theorem}

What remains it to describe the worst case performance of the (in expectation optimal) Randomized Rounding algorithm.

\begin{theorem} \label{theorem:5}
The objective value returned by Randomized Rounding algorithm is at most $\frac{2\ln(S)}{\xi^{\dag}}+3$ times worse than the optimal with high probability, where $\xi^{\dag}$ is the optimal objective value in the linear relaxed problem. 
\end{theorem}
\begin{proof}
The proof is similar to the previous theorems, yet the bound is tighter since we do not need to apply the Union Bound theorem. We begin by showing that:
\begin{equation}
\Pr [ \sum_{ u \in \mathcal{U} }   \widehat{y}_{lu} \geq (1+\epsilon)  \xi^{\dag}   ]  \leq  \exp^{\frac{-\epsilon^2  \xi^{\dag}   }{2+\epsilon}} \label{eq:boundB}
\end{equation}
Since $\xi^{\dag} \leq \widehat{\xi} $ where $\widehat{\xi}$ is the optimal integer solution value, it also holds that:
\begin{equation}
\Pr [ \sum_{ u \in \mathcal{U} }   \widehat{y}_{lu} \geq (1+\epsilon)  \widehat{\xi}   ]  \leq  \exp^{\frac{-\epsilon^2  \xi^{\dag}   }{2+\epsilon}} \label{eq:boundB}
\end{equation}
Next, we upper bound the right hand side of the above inequality by $1/S^2$. In order for this to be true, the $\epsilon$ value must satisfy the following condition:
\begin{equation}
\epsilon \geq \frac{\ln(S)}{\xi^{\dag}}  +\sqrt{   \frac{\ln^2(S)}{\xi^{\dag 2}_n}  +  \frac{4\ln(S)}{\xi^{\dag}}     }
\end{equation}
The above condition holds if we pick:
\begin{equation}
\epsilon= \frac{2\ln(S)}{\xi^{\dag}} +2
\end{equation}
since, in practice, the requests will be more than the services ($\xi^{\dag} \geq \ln(S)$). Thus, with high probability, performance will be at most $1+\epsilon = \frac{ 2 \ln (S)}{\xi^{\dag}} +3 $ times worse than optimal.
\end{proof}

The factors in Theorems \ref{theorem:1}-\ref{theorem:5} are \emph{bi-criteria approximations} with respect to both the objective value and capacity  constraints. In many practical scenarios, these factors are constant. For example, consider a system with thousands of users generating requests for services in a library of size $S=1,000$. Each BS  can process up to a thousand 
requests ($C_n = 1,000$) and the minimum computation capacity utilization is $40\%$ ($\lambda^{\dag}=400$). Then, the bi-criteria approximation factor becomes $\frac{3\ln(1000)}{400}+4 \approx 4.05$.

\section{Extension and Practical Cases} \label{section:extension}

In this section, we discuss how to handle changes in the user demand. Besides, we describe how to make the solution of the Randomized Rounding algorithm satisfy the capacity constraints, thereby making the algorithm more practical.

\subsection{Handling user demand changes}

The service placement and request routing decisions are taken for a certain time period during which the demand is fixed and predicted. The demand, however, may change over time, e.g., after a few hours or even at a faster timescale depending on the scenario. The MEC operator will have to repeatedly predict the new demand for the next time period and \emph{adapt} the service placement and request routing decisions accordingly. For example, the MEC operator should replace services that are no longer popular with other services that recently gained popularity.

The adaptation of the service placement is not without cost. In fact, replacing previously placed services with new ones would require from the BSs to download non-trivial amounts of data from the cloud through their backhaul links. This operation creates overheads which, depending on the timescale, can be significant and therefore should be avoided.


The Randomized Rounding algorithm can be extended to become aware of the service placement adaptation costs. To carry this out, we add a new constraint into the JSPRR problem. This constraint upper bounds by a constant $D$ the total amount of data associated with the replaced services:
\begin{align}
\sum_{n\in\mathcal{N}} \sum_{s\in\mathcal{S}}  x_{ns} (1 - x^{p}_{ns}) r_s \leq D \label{eq:adaptation}
\end{align}
where $x^p_{ns}$ is the placement solution in the previous time period. Here, placing a service $s$ at a BS $n$ ($x_{ns}=1$) adds $r_s$ to the adaptation cost if and only if that service was not placed in the previous time period ($x^{p}_{ns}=0$).

We note that all the presented lemmas and theorems still hold as they do not depend on the presence of constraint (\ref{eq:adaptation}). What remains to analyze is how likely is for the rounded solution $\widehat{\bm{x}}$ returned by the algorithm to violate constraint (\ref{eq:adaptation}). This is described in the following theorem.

\begin{theorem} \label{theorem:6}
The total amount of data associated with service placement adaptation will not exceed the upper bound $D$ by a factor of $\frac{ 2 \ln (S)}{D} +3 $ with high probability.
\end{theorem}
\begin{proof}
The proof is similar to the previous theorems. The Chernoff Bound is applied for the sum of random variables $\{x_{ns}(1 - x^{p}_{ns}) r_s\}$ the expected total value of which is $D$.
\end{proof}

\subsection{Constructing a feasible solution}

As the Randomized Rounding algorith may violate the storage capacities of the BSs by a factor of $3\ln(S)/R_n+4$, the MEC operator may not be able to store all the services required to ensure the performance guarantee of the algorithm. Similarly, the service placement may violate the limit of allowable adaptations $D$, while the request routing may overwhelm the computation and bandwidth capacities. To respond to such cases, the operator needs to convert the bi-criteria solution into a \emph{feasible} solution, i.e., a solution that satisfies constraints (\ref{eq:storage}), (\ref{eq:computation}), (\ref{eq:bandwidth-up}), (\ref{eq:bandwidth-down}) and (\ref{eq:adaptation}).

\begin{figure}[t]
	\begin{center}
		\includegraphics[scale=0.25]{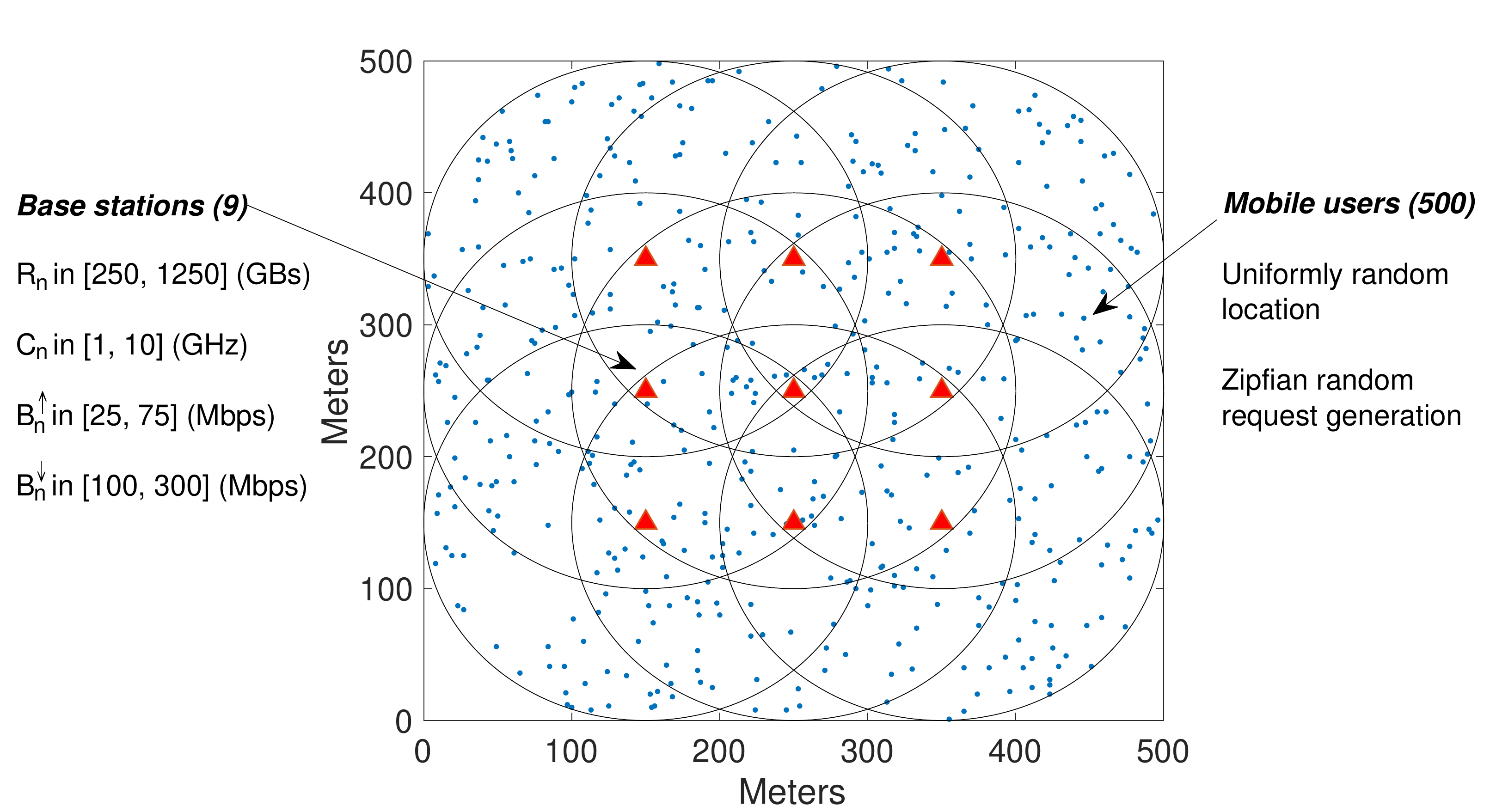}
		\caption{Evaluation setup.}
		\label{fig:setup}
	\end{center}
\vspace{-2mm}
\end{figure}

\begin{figure*}[t]
	\centering
	\subfloat[Impact of storage capacity.]{
		\includegraphics[width =4.3cm, height=4.1cm]{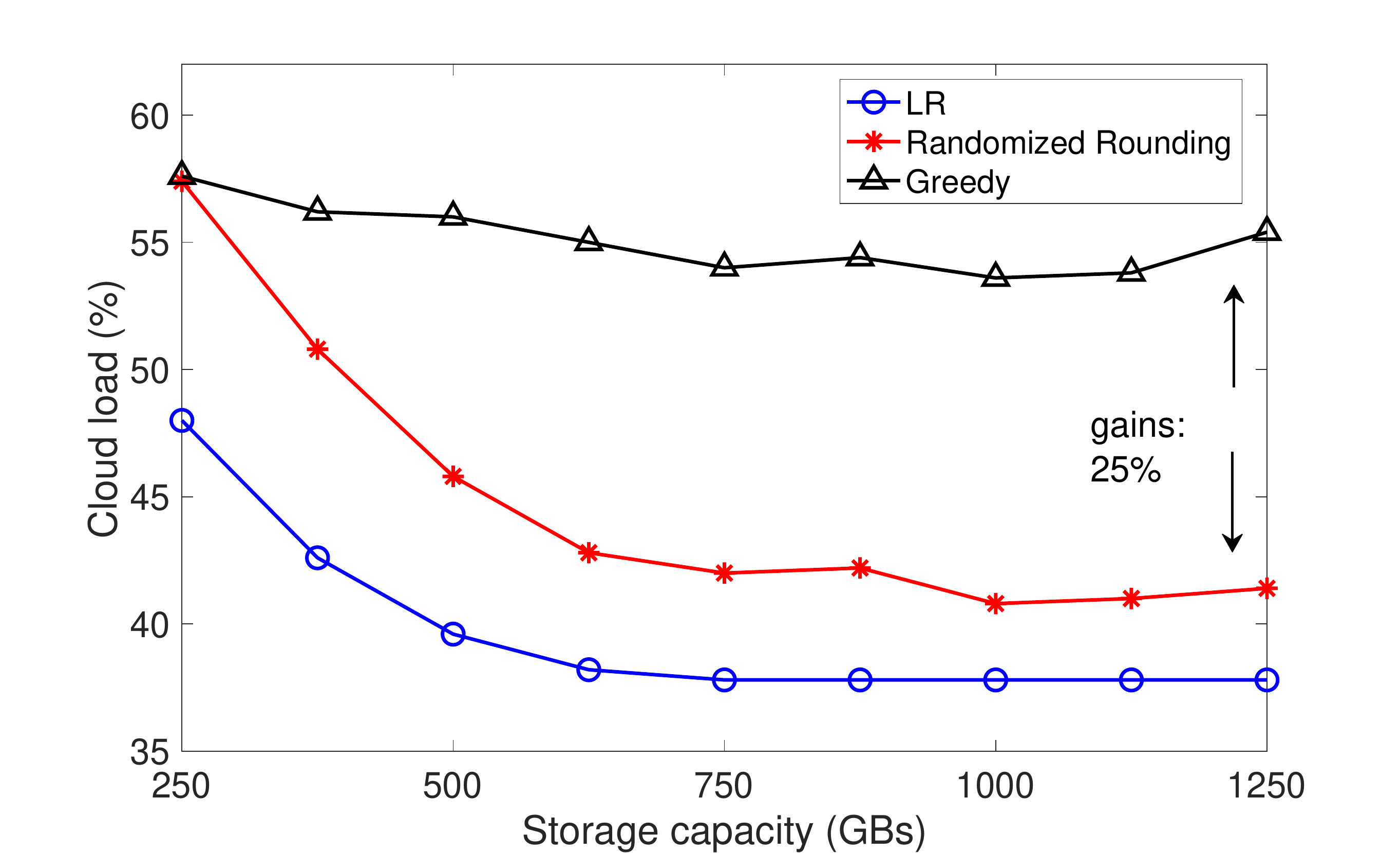}
		\label{fig:storage}
	}
	\subfloat[Impact of computation capacity.]{
		\includegraphics[width =4.3cm, height=4.1cm]{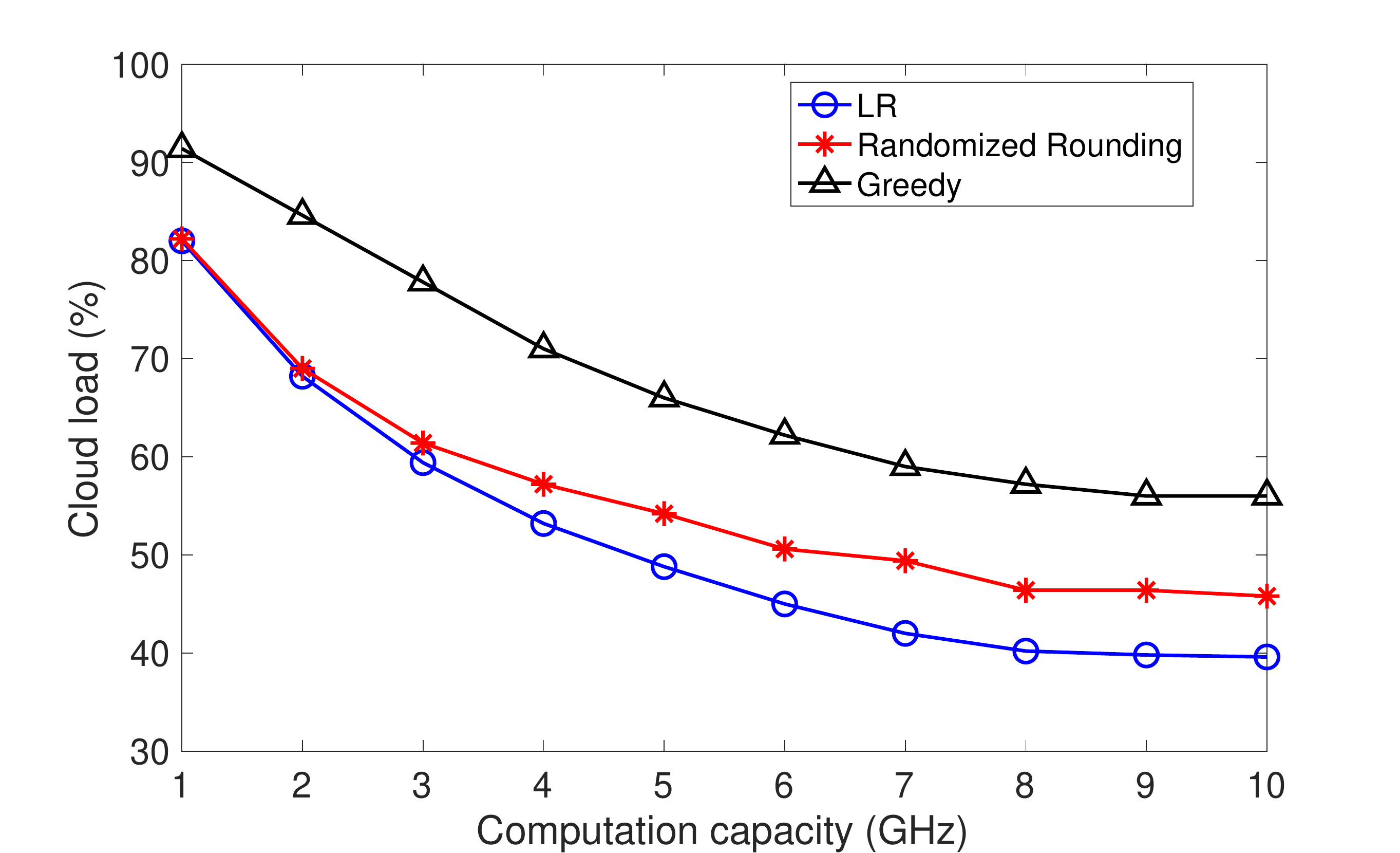} 		
		\label{fig:computation}
	}
	\subfloat[Impact of bandwidth capacities.]{
	\includegraphics[width =4.0cm, height=4.1cm]{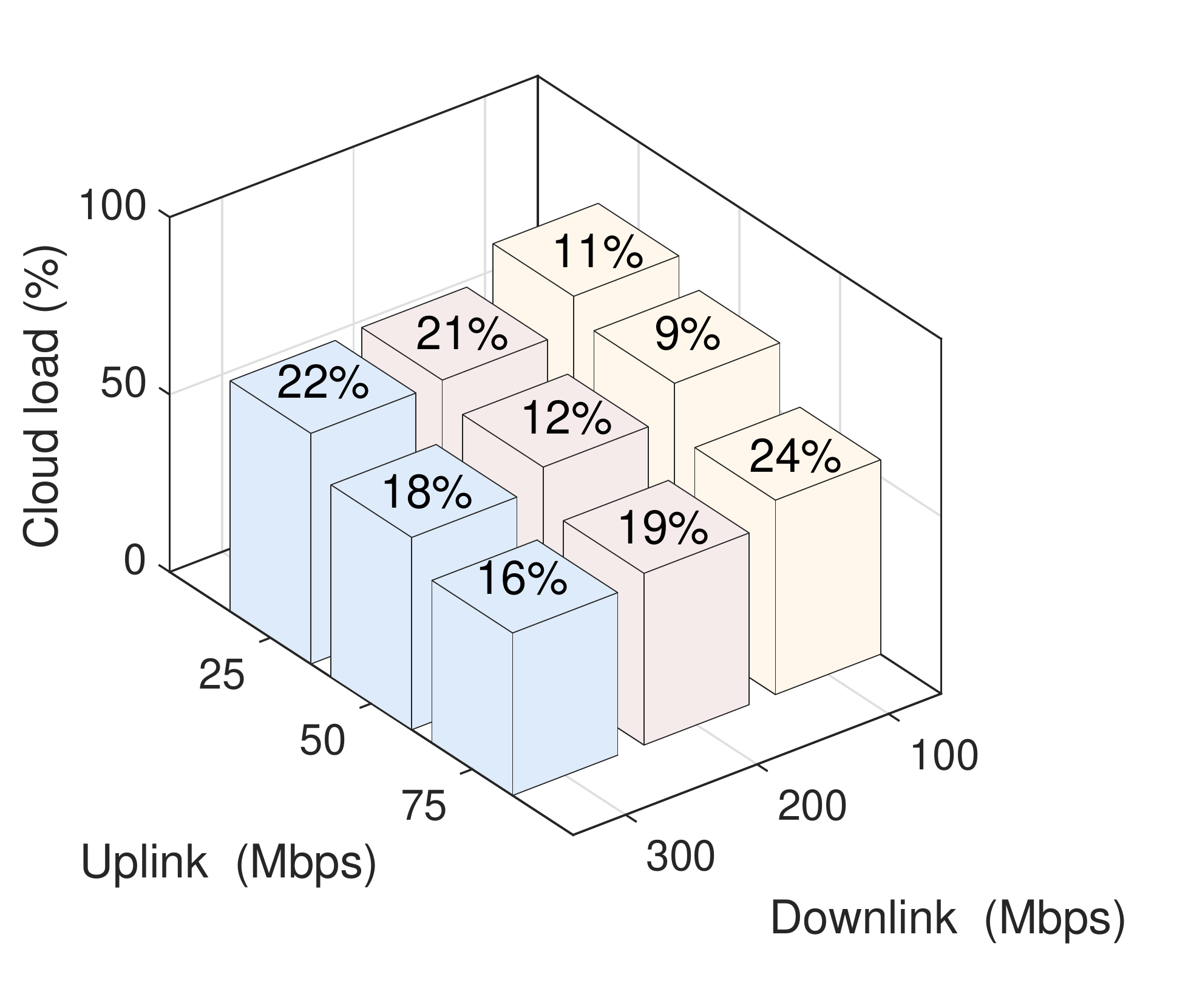} 		
	\label{fig:bandwidth}
	}
    \subfloat[Resource utilization per type and per BS.]{
	\includegraphics[width =4.8cm, height=4.1cm]{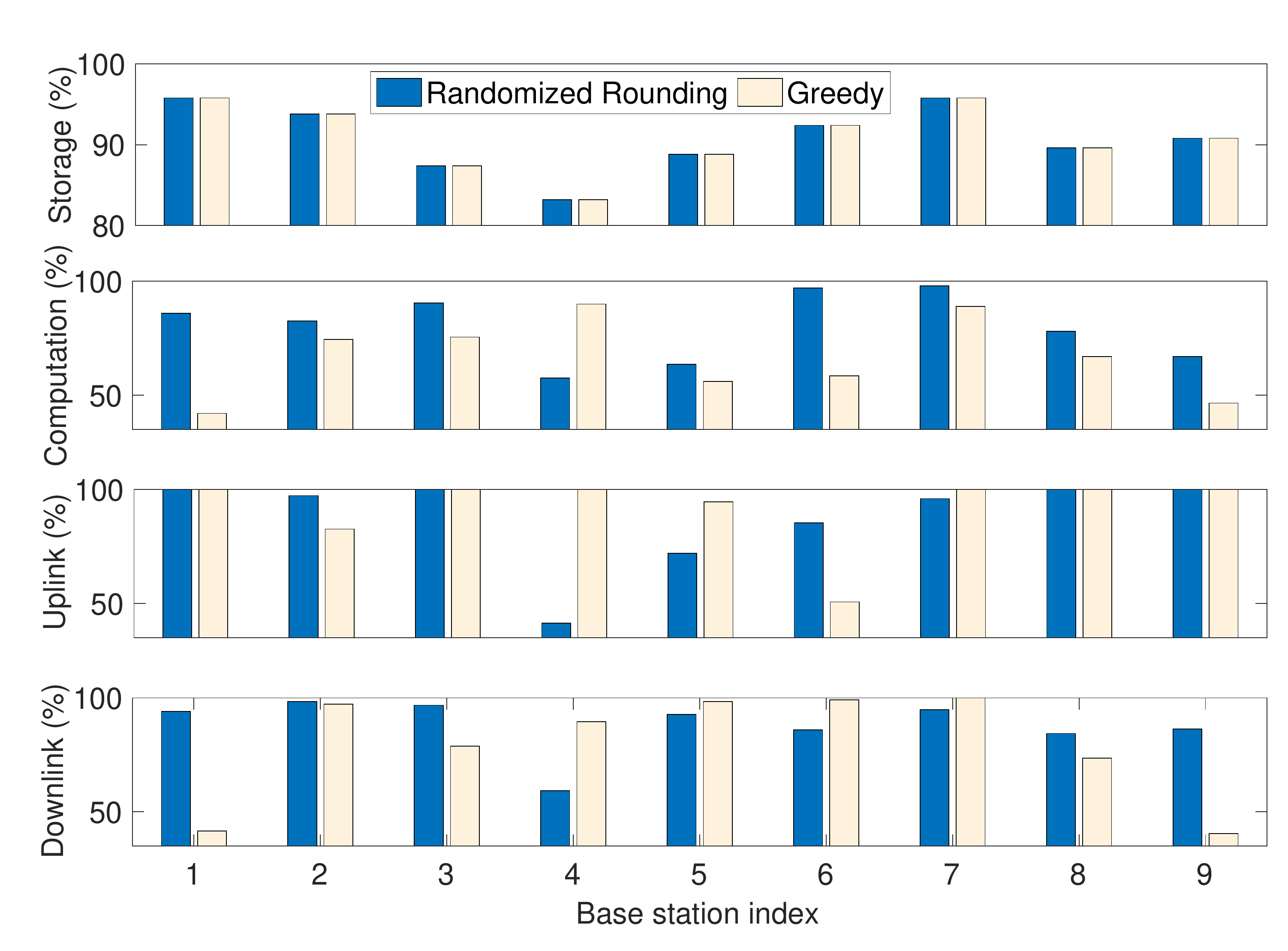} 		
	\label{fig:load}
	}
	\caption{Cloud load for different (a) storage, (b) computation and (c) bandwidth capacities of BSs. (d) Utilization of resources.}
	\label{fig:simulation}
\end{figure*}

To obtain such a solution, we start with the service placement $\widehat{\bm{x}}$ outputted by the Randomized Rounding algorithm. Then, we iteratively perform the removal of a service from a BS that yields the minimum cloud load increment. When a service is removed from a BS, the user requests for that service previously routed to that BS are now re-directed to other nearby BSs with available bandwidth and computation (if any), otherwise to the cloud. The procedure ends when constraints (\ref{eq:storage}) and (\ref{eq:adaptation}) are satisfied. To satisfy the remaining (computation and bandwidth) constraints we perform one more step. That is, we iteratively re-direct a user request from an overloaded BS to another BS with available resources (if any) or the cloud, until  there are not any overloaded BSs.


We need to emphasize that the above process may deteriorate the quality of the solution. However, as we show numerically in the next section, the obtained solution operates very close to the optimal in realistic settings.

\section{Evaluation Results} \label{section:evaluation}

In this section, we carry out evaluations to show the performance of the proposed Randomized Rounding algorithm. We consider a similar setup as in the previous work \cite{mec-storage}, depicted in Figure \ref{fig:setup}. Here, $N=9$ base stations (BSs) are regularly deployed on a grid network inside a $500$m$\times500$m area. $U=500$ mobile users are distributed uniformly at random over the BS coverage regions (each of 150m radius). Each user requests one service drawn from a library of $S=100$ services. The service popularity follows the Zipf distribution with shape parameter $0.8$, which is a common assumption for several types of services such as video streaming.

For each BS $n$, we set the storage capacity to $R_n=500$ GBs, the computation capacity to $C_n=10$ GHz and the uplink (downlink) bandwidth capacity to $B^{\uparrow}_n=75$ ($B^{\downarrow}_n=250$) Mbps. Yet, all these values are varied during the evaluations. For each service $s$, we set the occupied storage $r_s$ randomly within $[20, 100]$ GBs. The required computation per request $c_s$ takes value within $[0.1, 0.5]$ GHz. The required uplink (downlink) bandwidth per request $b^{\uparrow}_s$ ($b^{\downarrow}_s$) takes value within $[1,5]$ ($[1,20]$) Mbps.
We compare our algorithm with two baseline methods.
\begin{enumerate}
\item \emph{Linear-Relaxation (LR):} The optimal (fractional) solution to the linear relaxation of JSPRR problem. This solution is found by running a linear solver and provides a lower bound to the optimal integer solution value.
\item \emph{Greedy~\cite{femtocaching}:} Iteratively, places a service to a BS cache that reduces cloud load the most, until all caches are filled. Each request is routed to the nearest BS with the service, neglecting computation and bandwidth.
\end{enumerate}

On the one hand, LR can be used as a benchmark to gauge the performance gap of our algorithm from optimal. On the other hand, it is well-known that Greedy algorithm achieves near-optimal performance for the traditional data placement (or caching) problem, leveraging its submodular property~\cite{femtocaching}. Therefore, a natural question to ask is whether the efficiency of Greedy is maintained or novel algorithms are needed when the placement of services with multidimensional resource requirements is considered. 

We first explore the impact of storage capacity $R_n$ $\forall n$ on the load of the centralized cloud. In Figure \ref{fig:storage}, $R_n$ spans a wide range of values, starting from 250GBs to 1250GBs. As expected, increasing storage capacities reduces cloud load for all the algorithms as more requests can be satisfied locally (offloaded) by the BSs. \emph{The proposed Randomized Rounding algorithm performs significantly better than Greedy} with the gains increasing as the storage capacity increases (up to $25\%$ for $R_n=1250$GBs). At the same time, the gap from LR, and hence optimal, vanishes as $R_n$ increases (below $10\%$ for $R_n \geq 1000$GBs), which is consistent with the approximation factor expression in Theorem \ref{theorem:1}.

Next, we show the impact of computation capacity $C_n$ in Figure \ref{fig:computation}. While the cloud load reduces with $C_n$ for all the algorithms, Randomized Rounding performs consistently better than Greedy and very close to LR. Especially when $C_n$ is lower or equal to $3$GHz, the gap from LR is less than $3\%$. Similarly,  Figure \ref{fig:bandwidth} depicts the cloud load for different combinations of uplink ($B^{\uparrow}_n$) and downlink ($B^{\downarrow}_n$) bandwidth capacities. While the cloud load reduces with each of the $B^{\uparrow}_n$ and $B^{\downarrow}_n$ values, gains between $9\%$ and $24\%$ over Greedy are achieved (as shown in the bar labels).

Finally, we take a closer look into the utilization of BS resources when the Randomized Rounding and Greedy algorithms are used. The four subplots in Figure \ref{fig:load} show the resource utilization for each of the four resource types (storage, computation, uplink and downlink bandwidth). We observe that both algorithms  utilize most of the available storage resources ($90\%$ or more for most BSs). Yet, Randomized Rounding manages to utilize more computation resources for $8$ out of the $9$ BSs. This in turn will facilitate the offloading of more requests to the BSs while spending roughly the same or even slightly less bandwidth resources than Greedy.
\section{Conclusion} \label{section:conclusion}

In this paper, we studied joint service placement and request routing in MEC-enabled multi-cell networks with multidimensional (storage, computation and communication) constraints. Using a randomized rounding technique, we proposed an algorithm that achieves provably close-to-optimal performance, which, to the best of our knowledge, is the first approximation for this problem. This result can be of value in other research areas (e.g., data and middlebox placement). Interesting directions for future work include studying the coordination between BSs through backhaul links as well as the generalization of our model to services with multiple (chained) functions \cite{infocom17-jaime}.



\end{document}